\documentstyle{mn}
\input{psfig}

\newcommand{\mJybeam}{mJy beam$^{-1}$}
\newcommand{\whz}{W Hz$^{-1}$}

\newcommand{\etal}{{et al.}}

\begin{document}

\title{Radio jet interactions in the radio galaxy PKS~2152-699}

\author[R.A.E.  Fosbury et al.]{R.A.E.  Fosbury$^1$\thanks{Email: rfosbury@eso.org}, 
R. Morganti$^{2,3}$, W. Wilson$^2$, R.D. Ekers$^2$,
\newauthor S. di Serego Alighieri$^4$, C.N. Tadhunter$^5$
\\
$^1$ Space Telescope-European Coordinating Facility, D-85748 Garching bei
M\"unchen, Germany \\
$^2$ CSIRO, Australia Telescope National Facility, PO Box 76, Epping, NSW
2121, Australia\\
$^3$ Istituto di Radioastronomia, CNR, via Gobetti 101, 40129 Bologna,
Italy \\
$^4$ Osservatorio Astrofisico di Arcetri, largo E.Fermi 5, I-50125, Firenze,
Italy \\
$^5$ Department of Physics, University of Sheffield, Sheffield S3 7RH, England \\
}

\date{Accepted~~, Received~~}

\maketitle

\begin{abstract} We present radio observations of the radio galaxy
PKS~2152-699 obtained with the Australia Telescope Compact Array
(ATCA).  The much higher resolution and s/n of the new radio maps
reveals the presence of a bright radio component about 10 arcsec NE of
the nucleus. This lies close to the highly ionized cloud previously
studied in the optical and here shown in a broadband red snapshot image
with the HST PC~2. It suggests that PKS~2152-699 may be a jet/cloud
interaction similar to 3C277.3. This could cause the change in the
position angle (of $\sim 20^\circ$) of the radio emission from the inner
to the outer regions.  On the large scale, the source has
Fanaroff \& Riley type II morphology although the presence of the two hot-spots
in the centres of the lobes is unusual.  The northern lobe shows a particularly
relaxed structure while the southern one has an edge-brightened,
arc-like structure.

\end{abstract}


\section {Introduction}

Extended-emission line regions (EELR) are a common feature of powerful
radio galaxies. In the high-redshift objects, these regions display a
particular range of phenomena which are of especial interest for studies
of the properties of AGN and the formation and evolution of their host
galaxies. Extended line and rest-frame ultraviolet continuum emission
appears, above reshifts of around 0.8, to be aligned with the extended
double radio structures (McCarthy \etal\ 1987, Chambers, Miley \& van
Breugel 1987) although the radio and optical emitting regions are not
necessarily co-extensive. In some cases, however, there is a close
correlation between the radio emission and the EELR over distances of
tens of kiloparsecs and extreme kinematic components are observed
associated with the radio emission (e.g.,  McCarthy \etal\ 1987,
McCarthy 1993), suggesting a continuing interaction
between the radio plasma and the ISM.

In an effort to understand the high redshift phenomena in more detail, we are
studying selected low redshift objects which exhibit some of these
characteristics.  The correlation between EELR and radio emission is not so
common at low $z$.  Nevertheless, a number of cases of regions of ionized gas
found coincident or close to a radio feature --- either jet or lobe --- are known
and, in some cases, well studied (e.g., NGC~7385, Simkin \& Ekers 1979; 3C171,
Heckman, van Breugel \& Miley 1984; 3C277.3, van Breugel \etal\ 1985a;
Minkowski's object, van Breugel \etal\ 1985b; 4C~29.30, van Breugel \etal\
1986; Centaurus ~A, Morganti \etal\ 1990; 3C285, van Breugel \& Dey 1993; PKS~2250-41, 
Clark \etal\ 1997 and
PKS~1932-43, Villar-Martin \etal\ in prep.).  Some of them are believed to
represent an interaction between the radio plasma and the ISM: a `jet/cloud'
interaction.

PKS~2152-699 is a powerful, nearby radio galaxy ($z=0.0282$, 1 arcsec = 0.8
kpc for H$_\circ$=50 km s$^{-1}$ Mpc$^{-1}$) with a cloud of highly ionized
gas observed at about 10 arcsec ($\sim$8 kpc) from the nucleus.  The very
highly ionized cloud (HIC), with ions up to Fe$^{9+}$ and a blue optical
continuum, was first studied in detail by Tadhunter \etal\ (1987, 1988).
Although the HIC has a velocity very similar to systemic, broad, blue wings in the
[O~III] emission lines (up to 3000 km/s from the systemic velocity of the HIC
and galaxy) have been observed.  The presence of some gas at high velocities
led Tadhunter \etal\ (1988) to conclude that the cloud could possibly be
related to interaction between the radio plasma and the ISM.

However, later work (di Serego Alighieri \etal\ 1988; Fosbury \etal\
1990) showed that the HIC has a very blue optical to near-UV continuum
with 10\% linear optical polarization and the E-vector perpendicular to
the position angle of the nucleus-HIC axis.  The existence of the blue,
polarized continuum led di Serego Alighieri et al.  (1988) to suggest
that this resulted from the scattering by dust in the cloud of beamed
radiation from the nucleus. In this picture, the source of ionization
for the cloud is UV radiation from the nucleus. The presence of transverse 
ionization gradients also lent support to the beamed illumination model 
(Tadhunter \etal\ 1988).

The quality of the radio maps available until now (Christiansen \etal\
1977, Tadhunter \etal\ 1988, Jones \& McAdam 1992, Fosbury \etal\ 1990)
does not allow a detailed investigation of the radio emission in the
region of the HIC. To search for clear evidence of a jet/cloud
interaction in PKS~2152-699, we have made new radio observations with
the Australia Telescope Compact Array (ATCA) at both 3 and 6~cm
wavelength. In presenting the radio data and comparing with the optical
morphology, we use a broadband red image of the galaxy taken with the
HST PC~2  available from the public archive. VLBI observations of the nuclear
region were presented by Tingay \etal\ (1996).  

Subsequent papers will report a series of infrared, optical, ultraviolet and
X-ray observations which allow us to study the broad-band spectral energy
distribution of the nucleus and separate components within the interaction
site.

\section{Observations}

The observations reported in this paper were made on the 19 and 23 January, 21
March and 1 April 1992 using the four 6-km configurations available with ATCA
and the standard continuum correlator setup with a bandwidth of 128 MHz and 32
channels.  The two simultanous frequencies were set to 4.74 and 8.64 GHz.  The
source was observed for 12h in each configuration.  The data were calibrated by
using the MIRIAD package (Sault, Teuben \& Wright 1995), which is necessary for
the calibration of the polarization of ATCA data.  The flux scale is based on
the recent compilation of mesurements of the primary calibrator PKS~1934-638 by
Reynolds (1996) which corresponds to 5.91 Jy at 4.7 GHz and 2.84 at 8.6 GHz.
This differs from the previous calibration in use at the Compact Array by
(new-old) -8.1\% at 4.7 GHz and +9.8\% at 8.6~GHz.

At 4.7~GHz we have made images with both uniform weight and using the
``robust'' parameter equal to 0.5.  The former gives the full resolution of
$1.39 \times 1.44$ (PA = $11.7^\circ$) while the latter gives a beam of $2.3
\times 2.1$ arcsec (PA = $58^\circ$).  Using the ``robust'' parameter we were
able to give more weight to the short baselines and therefore the extended
low-brightness emission is better imaged.  The total intensity map at 4.7~GHz
at lower resolution is shown in Fig.~1.  At 8.6~GHz, the data have a full
resolution of $0.8 \times 0.9$ arcsec (PA = $ - 26^\circ$). The rms noise of 
the total intensity maps is about 0.80
\mJybeam\ in the 4.7~GHz map and 0.65 \mJybeam\ in the 8.6~GHz map.

\begin{figure}
\centerline{\psfig{figure=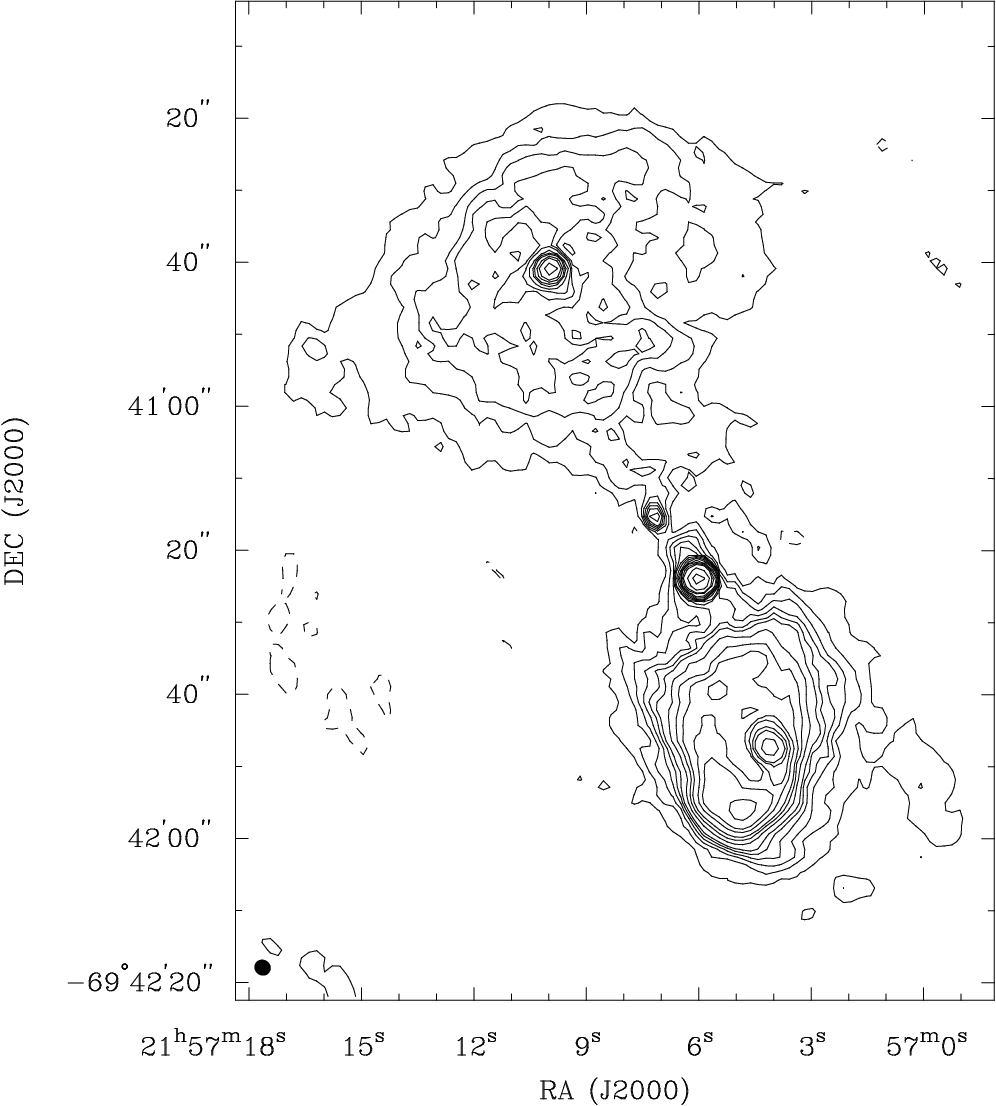,width=9cm,angle=-90}}
\noindent{\bf Figure 1.}  ATCA 4.8~GHz image of PKS~2152-699. The
contour levels are --1, 1, 2, 3, 4, 5, 6, 8, 10, 12, 16, 20, 30, 40, 80, 160
$\times$ 3 mJy
beam$^{-1}$.
\end{figure}

The total fluxes derived from these synthesis observations are given in
Table~1.  At 8.6~GHz, it appears to be lower than single dish observations
(6.22 Jy at 8.4 GHz from Wright \& Otrupcek 1990), most likely due to missing short
spacings. 

\begin{table}
\begin{center}
\caption{Radio Properties of PKS~2152-699}
\begin{tabular}{@{}lcccc}
         &            &           &            &  \\
Region   &  I$^{a}_{\rm 4.7GHz}$ &    m$^a_{\rm 4.7 GHz}$  & I$^{b}_{\rm
8.6GHz}$ & $\alpha^{4.7}_{8.6}$     \\
         &    Jy      &     \%       & Jy    &  \\
Total    &   9.94    &    24.0  &  3.80  &  -- \\
Core     &   0.77    &     2.0  &  0.81  &  $+0.10$   \\
HIC      &   0.04    &    21.9  &  0.017 &  $-1.10$ \\
N lobe    &  3.61    &    26.4  &   --  &  --  \\
N hot spot & 0.21     &   14.6  &  0.09  &  $-0.87$  \\
S lobe     & 5.41    &    21.2  &  2.30  &   --  \\
S hot spot & 0.56   &     13.8  &  0.25  &  $-0.92$  \\
\end{tabular}
\end{center}
a) beam size 2.1$\times$2.3 arcsec \\
b) beam size 1.5$\times$1.6 arcsec \\
\end{table}

We have also obtained the images of the Stokes parameters ($Q, U$), the
polarized intensity image ($P=(Q^2+U^2)^{1/2}$) and position-angle image
($\chi=0.5 {\rm arctan} (U/Q)$).  Here we present these images only for the
data at 4.7~GHz because at 8.6~GHz the polarization maps are not completely
reliable, probably due to instrumental polarization affecting the data in the
earliest days of ATCA when the data presented here were
collected.  The rms noise of the $Q$ and $U$ maps at 4.7 GHz is about 0.17
\mJybeam.

The polarized intensity and the fractional polarization ($m=P/I$) were
estimated only for the pixels for which $P>5\sigma_{QU}$.  Fig.~2 shows
a greyscale image of the polarized intensity while Fig.~3 shows contours
of the total intensity with superimposed vectors whose length is
proportional to the fractional polarization and whose position angle is
that of the electric field.  The mean fractional polarization of the
different regions is given in Table~1.

\begin{figure}
\centerline{\psfig{figure=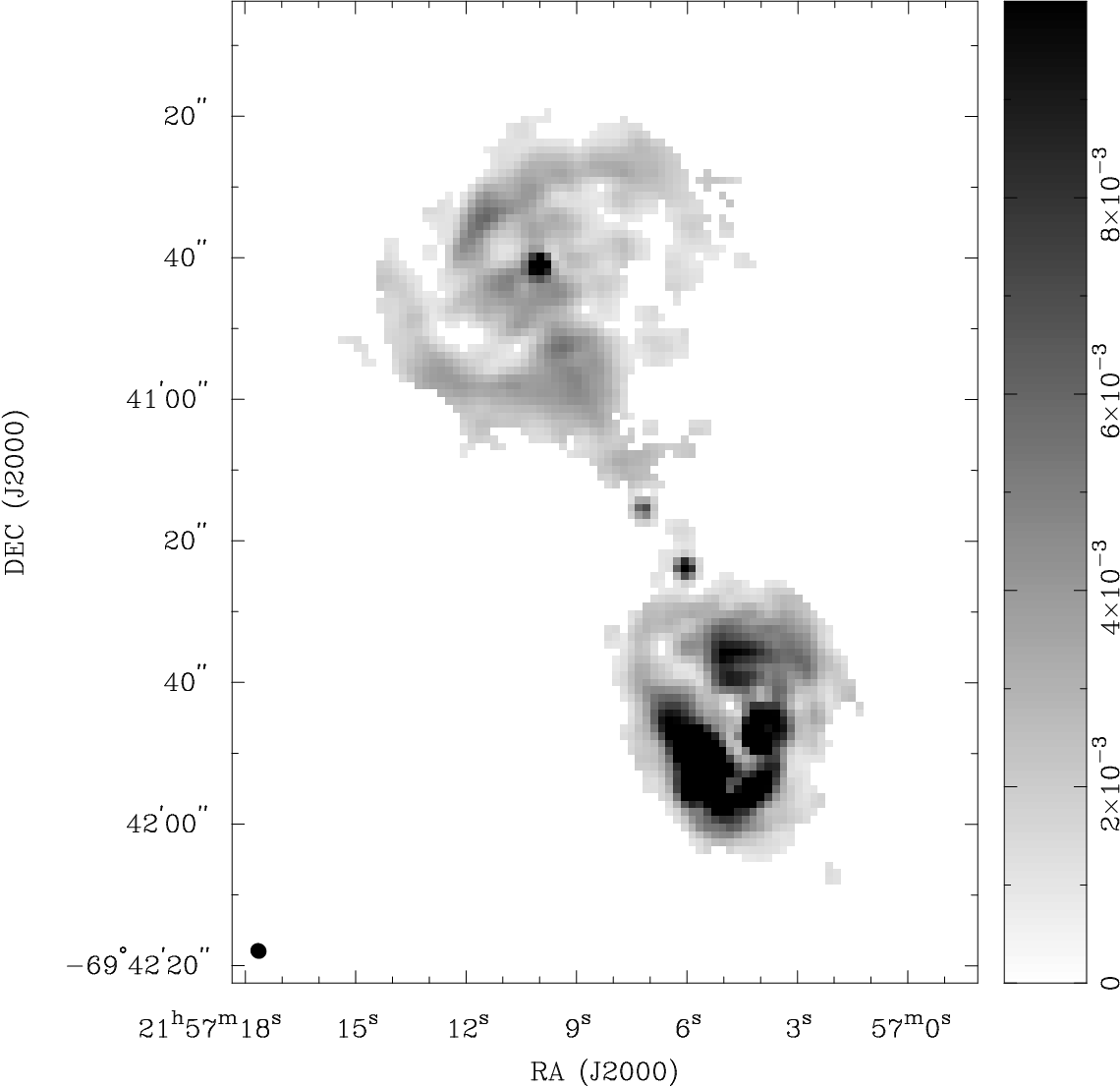,width=9cm,angle=-90}}
\noindent{\bf Figure 2.} Greyscale image of the polarized intensity of
PKS2152-69 at 6cm. The range is between 0 and 10 mJy.
\end{figure}

\begin{figure}
\centerline{\psfig{figure=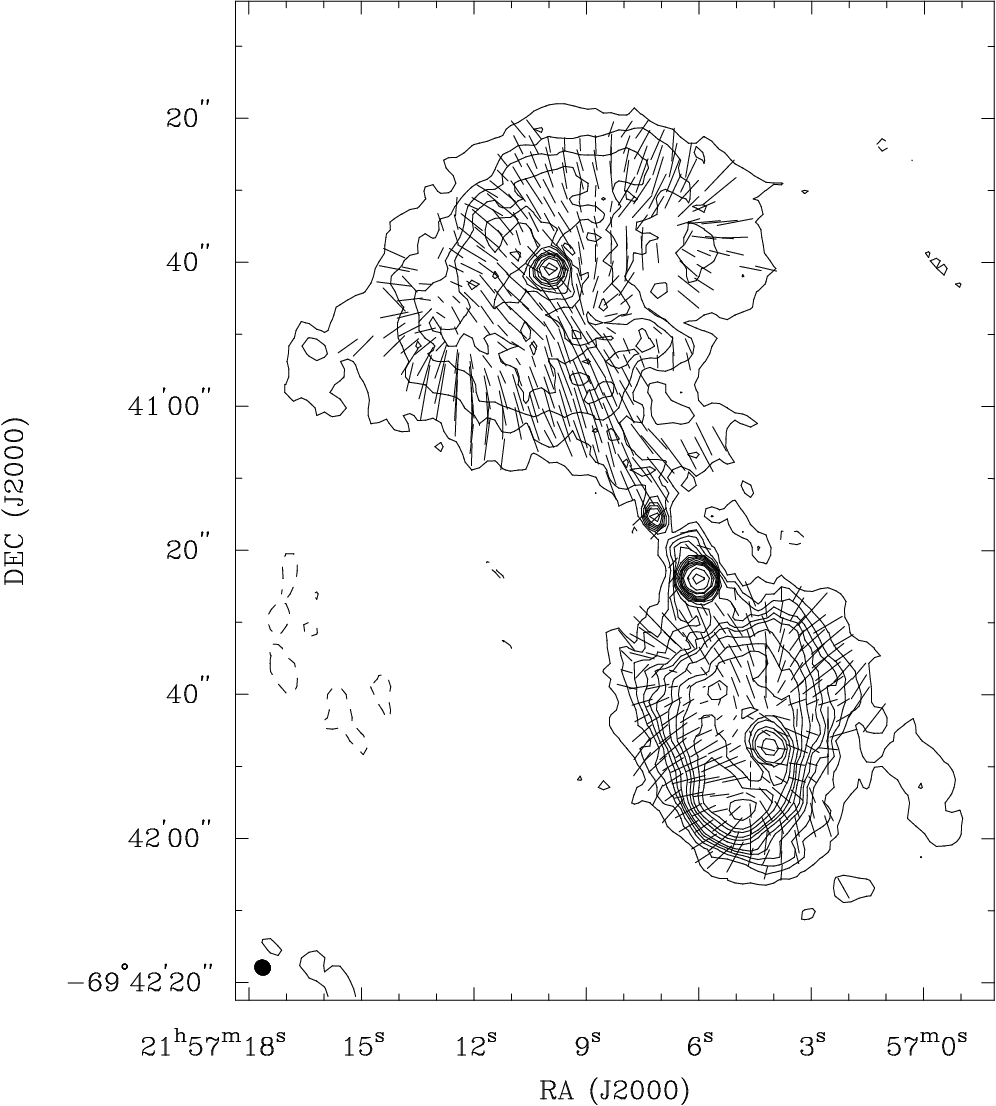,width=8cm,angle=-90}}
\noindent{\bf Figure 3.} Contours (as in Fig.~1)
of the total intensity at 4.8~GHz image of PKS~2152-699
with superimposed vectors whose length is proportional to the
fractional polarization and whose position angle is that of the electric
field.
\end{figure}

In order to estimate the spectral index in some regions we have produced an 8.6~GHz map
degrading the resolution to match that of the 4.7~GHz (uniform weight).  The
spectral index $\alpha^{4.7}_{8.6}$ (defined as $S\propto \nu^{\alpha}$) was
estimated only from the regions with signal above $5\sigma_I$ in both maps.
The values of the spectral index  must be interpreted with caution because the
observations were not made 
with matched arrays: a more complete study of both polarization and spectral
index is in progress and will be presented in a forthcoming paper.

\subsection{Results}

Compared to the previous observations from the Molonglo
Synthesis Telescope (Tadhunter \etal\ 1988) and from ATCA (Fosbury \etal\
1990), the radio structure is now shown more clearly. An important
feature is the bright source $\sim 10$ arcsec NE of the nucleus, i.e.,
at the same distance and similar---but not identical---position angle
from the nucleus as the HIC.  This component, which we call RC, 
was not visible in the
previous low resolution radio images available for PKS~2152-699 although
it was seen with a lower significance in the early ATCA image by Fosbury
\etal\ (1990).

The new radio images show that PKS~2152-699, although strictly classified as a
Fanaroff \& Riley type II source on the basis of the ratio of hot-spot separation 
to total source length, it is unusual in showing the hot-spots near the centres 
of the lobes. This behaviour, which is intermediate between FR~I and II types, has 
been observed
before (e.g.  Capetti, Fanti \& Parma 1995) in objects with radio power on the
border between the two classes ($\log {\rm P} \sim 25.5$ \whz\ at
4.7~GHz).  The total radio power of PKS~2152-699 ($\log {\rm P} =
25.61$ \whz\ at 4.7~GHz) is indeed typical of the transition region
between FR~I and II galaxies.  For the objects studied by Capetti
\etal\ (1995), the peculiar morphology could be due to precession of
the central engine or perhaps by an externally-induced change in jet
direction.  The southern lobe of PKS~2152-699 shows an
edge-brightened, arc-like structure which suggests jet precession.
The bright spot then represents the primary hot-spot and the current
point of impact of the jet with the ISM while the arc-like bright
emission represents the previous impact points during the change of
direction of the jet (see also the simulations in Cox, Gull \& Scheuer 1991).
Something similar could also occur in the northern lobe with the radio
plasma far away from the hot-spot could represent the backflow from
previous, more distant, fading hot spots.  A similar scenario, with
the lobe considered as a relic from earlier jet activity, has been
suggested for 3C111 (Linfield \& Perley 1984).  Hot-spots in the
middle of the lobe can also be due to projection effects or the result
of an interaction moving the jet direction.  The spectral indices
observed in the hot-spots are unusually steep ($\alpha^{4.7}_{8.6}
\sim -0.9$) which could be due to the result of a spectral break
between these two frequencies. This has been observed in the knots of
3C277.3, an object with many characteristics in common with
PKS~2152-69.

The radio emission in PKS~2152-699 is asymmetric both in flux and
core/lobe distance. The southern lobe extends for
approximately 30~arcsec (24~kpc) while the northern extends to
45~arcsec (36~kpc).  We do not find clear evidence for a jet, but a
thin, low brightness emission bridge between the nucleus and the
northern lobe is visible at 4.7~GHz. Although very weak, the total
emission from the northern lobe shows substructure. This
substructure can be seen very clearly in polarized emission (Fig.~2).
Two arm-like structures are visible along its southern and northern
edges.  The northern lobe shows a somewhat higher fractional
polarization while the hotspots appear to be less polarized than the
rest of the lobes.  In the southern lobe, the polarization vectors
follow the edge-brightened arc-like structure (see Fig.~3), typical of
such sources.
The fractional polarization in the lobes is quite high although not
exceptional.

PKS~2152-699 has a prominent core (see Table~1).  High resolution
observations made in 1988-89 with the Parkes-Tidbinbilla interferometer
by Jones \etal\ (1994) give a core flux of 784 mJy at 2.3GHz
while more recent observations obtained in 1994 with the same instrument
give 583 mJy at the same frequency (Morganti
\etal\ 1997).  This may indicate some variability of the flux
density of the nucleus although new observations will be necessary to
confirm this. The core has an inverted spectral index, $\alpha \sim 0.10$.

The ratio $R$ between the core flux density and the extended flux
density derived at 4.7 GHz is $R \sim$0.06.  This is a typical value for
broad-line galaxies as PKS~2152-699 is classified by Tadhunter \etal\
(1988) while it is large compared to typical FR~II narrow-line radio
galaxies as discussed in Morganti \etal\ (1997) for data collected at
2.3~GHz.

\subsection{The nearby galaxy 2153-699}

PKS~2152-699 is known to be situated in a galaxy-poor environment (Tadhunter
\etal\ 1988).  However, the presence of a nearby ($\sim 3.8$ arcmin east)
strong radio source, PKS~2153-69, was already known from the model fitting to
low resolution data carried out by Ekers (1969).  This source is unresolved in
the maps of Christiansen \etal\ (1977) and Tadhunter \etal\ (1988) and has
been identified by Jones \& McAdam (1992) with a 20th magnitude galaxy with
unknown redshift but likely to be in the background given its magnitude.  With the
new data we have mapped the region corresponding to PKS~2153-699 and 
resolved the source into two lobes or tails.  Some trail of radio emission to
the west is also observed.  More faint structure may be missed because of the
attenuation of the primary beam.  In Fig.~4 the contours of the radio emission
are shown.  The cross indicates the position of the optical identification,
situated between the two lobes. 

\begin{figure}
\centerline{\psfig{figure=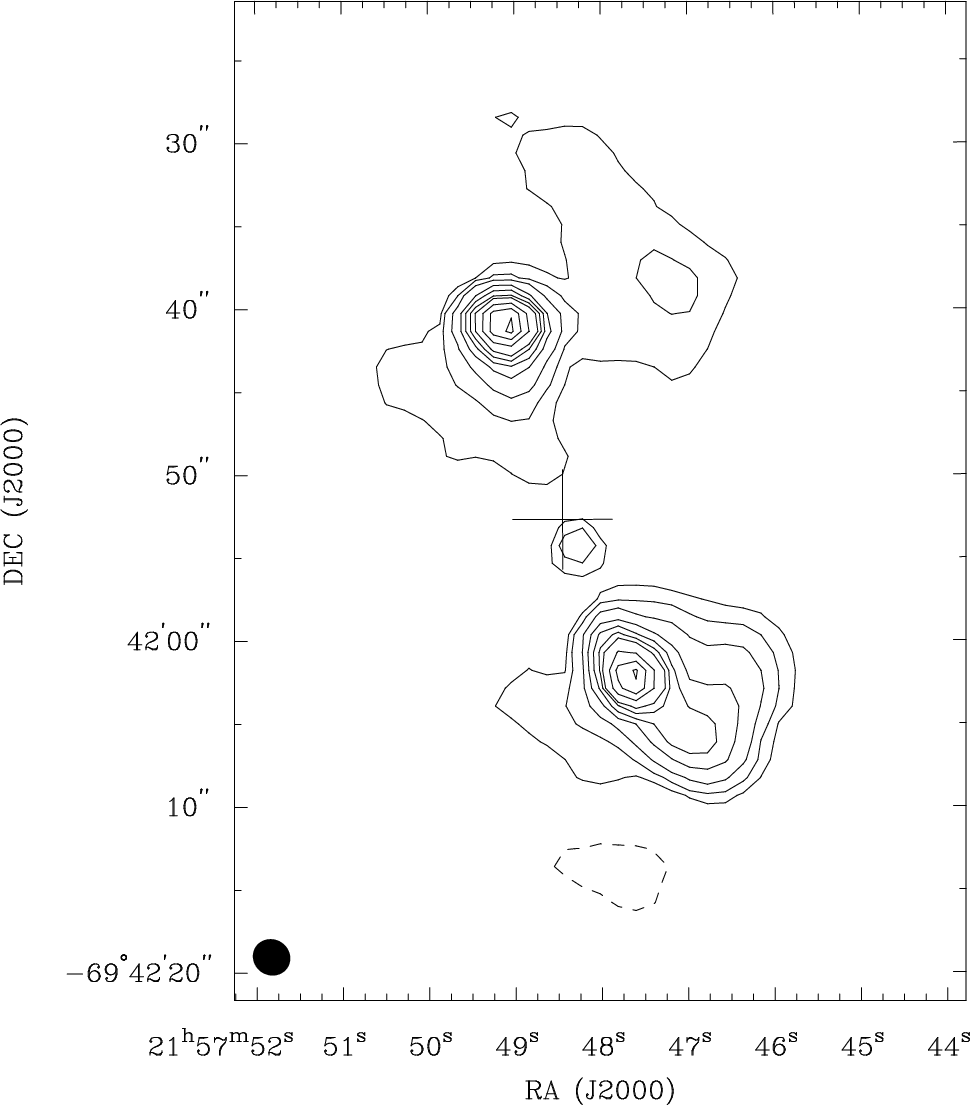,width=8cm,angle=-90}}
\noindent{\bf Figure 4.} Contours
of the total intensity at 4.8~GHz image of the nearby galaxy
PKS~2153-699. The contour levels are
-1, 1, 2, 4, 6, 8, 10, 12, 14, 16, 20, 30 $\times$ 2 mJy beam$^{-1}$
\end{figure}

\subsection{Correlation with the optical image}

A broad-band HST PC~2 image (F606W filter, single 500s exposure) of
PKS~2152-699 was retrieved from the public archive.  After interactively 
cleaning cosmic-ray
events, we have used this image to produce an overlay with the new
radio image at 4.7GHz.  The registration between the two images was carried
out by aligning the radio nucleus (RA(J2000) = 21 57 06.0, DEC(J2000) = -69 41
24.0) with the peak of the optical image ($\sigma \sim$ 0.2 arcsec).  The
overlay is shown in Fig.~5 while Fig.~6 presents a cartoon which shows more
clearly the spatial relationship between the radio, optical emission and dust
absorption components.  The HST image, which includes line (predominantly
[O~III], H$\alpha$ and [N~II]) emission and continuum, clearly shows
structural details of the HIC and several additional components which fall
along the same axis.  The position angle between the nucleus and the observed
radio component, RC, near the HIC is 34$^\circ$.  This is intermediate between that of the
large-scale radio structure (PA$\sim 23^\circ$ defined from the low resolution
MOST radio image, Tadhunter \etal\ 1988) and that from VLBI observations
presented by Tingay \etal\ (1996).  The VLBI observations reveal a core-jet
morphology on the parsec scale and the PA of this structure is $\sim 44^\circ
\pm 5^\circ$, close to that of the centroid of the optical HIC.  The overlay
shows that the HIC is situated to the east of RC which itself is coincident 
with a prominent part of the substructure of the cloud. This latter component was
identified as a red stellar object (C) in the line-free continuum observations
reported by di Serego Alighieri \etal\ (1988)
and is therefore predominantly continuum rather than line emission with a colour very
different from the rest of the HIC.  The dust
band crossing the galaxy in PA$\sim 110^\circ$ was inferred from the
groundbased images from the colour gradient measured across the galaxy (di
Serego Alighieri \etal\ 1988). 

\begin{figure*}
\vspace*{14 cm}
\noindent{\bf Figure 5} HST F606W image with radio 6cm contours superimposed. The HST
image has been manually cleaned of cosmic ray events and is shown on a logarithmic
scale. The radio contours are also logarithmic, ranging from 3 to 700 \mJybeam. The
radio and optical nuclei have been forced to coincide. The regions containing some of
the faint features are shown with a stretched intensity scale.
\end{figure*}

\begin{figure}
\centerline{\psfig{figure=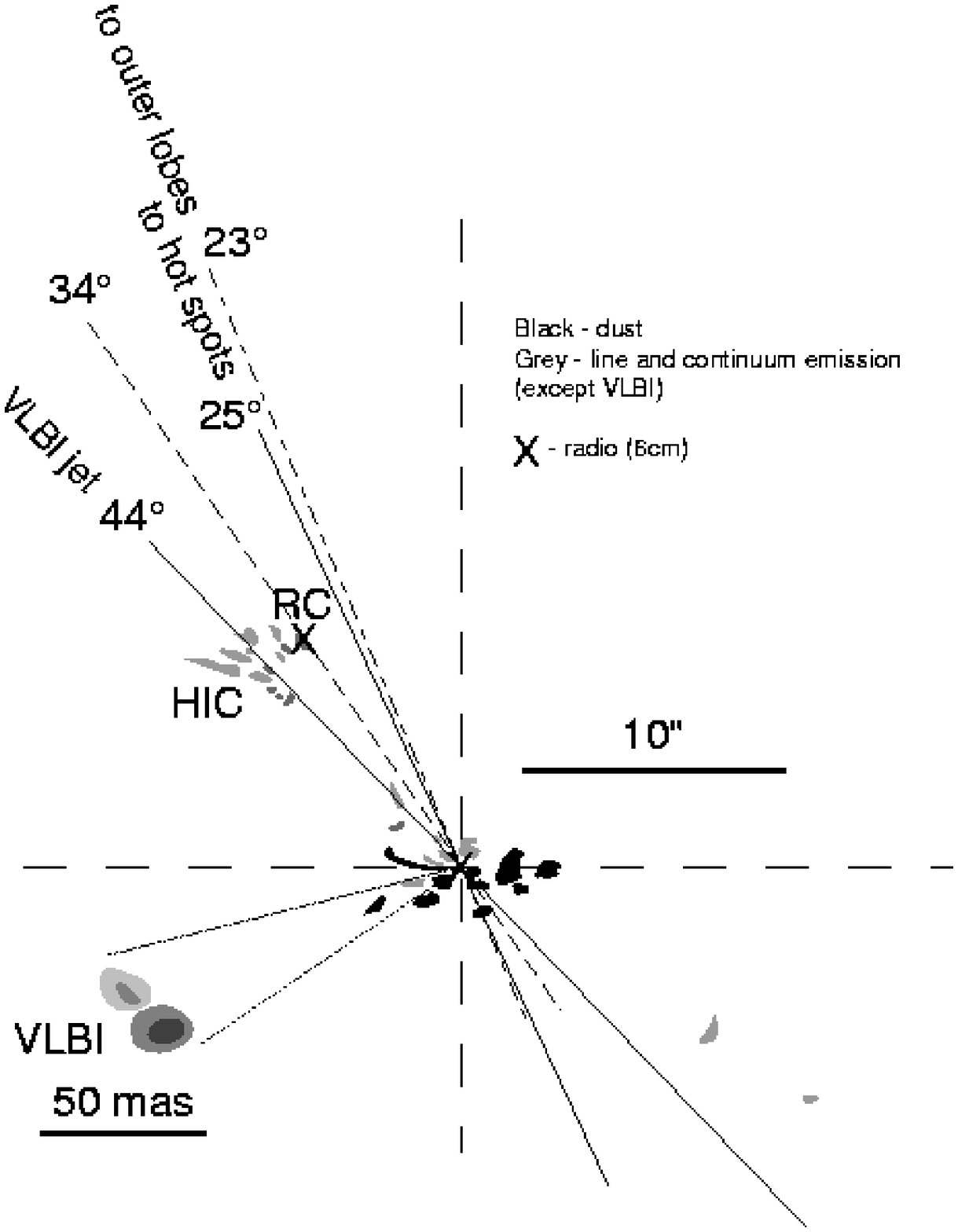,width=9cm,angle=0}}
\noindent{\bf Figure 6} Cartoon showing the different absorption and emission
components in PKS2152-699. The VLBI image is represented to a different scale
at the lower left.
\end{figure}

In addition to the emission components to the NE of the nucleus, which extend
to greater distance than shown in Fig.~5 (Tadhunter \etal\ 1988), the HST
image shows two faint regions to the SW at radial distances of approximately
11 and 16~arcsec in PA~237$^\circ$.  These were not detected in the
groundbased images.

There is also a bright `bow' structure, convex to the
nucleus, approximately 3 arcsec along PA$\sim57^\circ$ to the NE.  This is
diametrically opposite the faint SW structures.  This bow was seen as a
nuclear extension in the groudbased emission line imaging (see Fig.~5).

\section{Discussion}

The combination of radio and optical images of PKS~2152-699 allow us to make
the following statements:

\begin{enumerate}

\item{There is a radio component associated with the HIC --- although it
is not situated at the brightest optical location. The optical emission
coincident with this is red in colour and probably continuum rather than
line emission.}

\item{There is a progression in PA from the VLBI axis ( = the nucleus --
HIC axis) to the nucleus -- inner radio component to the large-scale
radio double axis (44 -- 34 -- 23$^\circ$).}

\item{There are a
number of distinct optical structures (of which the
HIC is the most prominent) lying close to the 44$^\circ$ PA axis. These
exist on both sides of the nucleus although they are more numerous and
brighter to the NE.}

\item{A dust lane crosses the nucleus in PA$\sim110^\circ$ with emission
along its northern edge. This could be the low ionization line emission reported by
Tadhunter \etal (1988) but we cannot exclude a continuum contribution.}

\item{The northern and southern lobes have a very different morphologies
--- the northern being diffuse with a central hotspot, the southern
brighter with a `trailed' hotspot.}

\end{enumerate}

The discovery of the bright radio component next to the HIC supports the
hypothesis of a jet/cloud interaction in PKS~2152-699 and it reinforces
the similarity between this object and 3C277.3, one of the best examples
of jet/cloud interaction (van Breugel \etal\ 1985a).  As pointed out by
Tadhunter \etal\ (1988), comparison of the two objects shows a
similarity in the emission-line spectra, the presence of the blue
optical continuum and the offset between the cloud/nucleus axis and the
radio axis. They also pointed out, however, that the cloud of ionized
gas in 3C277.3 shows a large velocity gradient while the HIC in
PKS~2152-699 shows broad, low intensity wings on the [O~III] lines 
extending to the blue. The
presence of extensive, chaotic dust lanes in the inner regions
of the galaxy suggests that the the radio jet is interacting with a
fragment of a merging galaxy.

We have estimated for PKS~2152-699 the minimum pressure associated with the
bright radio component to be $\sim 3 \times 10^{-10}$ dynes cm$^{-2}$.  We do
not have a good measure of the density of the warm line-emitting gas but if we
assume T$\approx$15000 K (Tadhunter \etal 1988) and $n_e \sim 200$ cm$^{-3}$
(a typical value for a bright extranuclear cloud) we find that, as in the case
of other jet/cloud interactions, the radio and line-emitting gas pressures are
quite comparable (Clark \& Tadhunter 1996) suggesting that the region of high
ionization has been compressed by the interaction with the radio plasma.  

The presence of EELR in high and intermediate redshift radio galaxies is often
associated with asymmetries in the radio morphology and polarization (Pedelty
\etal\ 1989, Liu \& Pooley 1991, Clark \etal\ 1997). In PKS~2152-699, however,
there is no strong polarization asymmetry and the stronger, closer 
(in the sky plane) radio lobe is
on the side opposite to the jet/cloud interaction. In contrast with the more
powerful radio galaxies at higher redshifts, where the polarization and
structural asymmetries appear to be due to large-scale ISM density variations
across the source, the interaction seen here may be a transitory phenomenon
without a major influence on the main radio lobe structure.

\section{Conclusions}

We have presented new radio maps of the radio galaxy PKS~2152-699
obtained with the ATCA.  These reveal a bright radio component next to the
cloud of very highly ionized gas, situated about 8~kpc from the nucleus.
The radio component is separated from the brightest region of the
optical cloud but coincides with source of red continuum
emission. This structure, together with the unusual morphology of the
northern radio lobe, suggests that a jet/cloud interaction has taken
place. Such interactions are relatively rare at low redshift but, when
they occur, do offer the opportunity for quantitative investigation of
the properties of the jets themselves. We are currently persuing such
studies using these and other new observational data on this source.

\section*{Acknowledgements}

This work is based on
observations with the Australia Telescope Compact Array (ATCA), which
is operated by the CSIRO Australia Telescope National Facility. It also uses
observations made with the NASA/ESA Hubble Space Telescope obtained from the data
archive at the ST-ECF.
RAEF is  affiliated to the
Astrophysics Division, Space Science Department, European Space Agency.
RM acknowledges support from the  DITAC International Science \& Technology
Program.

{}
\end{document}